\documentclass[aps,amssymb,orl,twocolumn]{revtex4}
\usepackage{amssymb}

\usepackage{graphicx}
\usepackage{amsmath}
\usepackage{times}

\begin{document}

\title{  Lifshitz-like transition and enhancement of correlations in a rotating
bosonic ring lattice}

\author{ Ana Maria Rey$^{1}$\footnote{Electronic address: arey@cfa.harvard.edu}, Keith Burnett $^{2}$ ,
Indubala I. Satija$^{3,4}$ and Charles W. Clark$^{3}$}
\affiliation{$^{1}$ Institute for Theoretical Atomic, Molecular and
Optical Physics, Cambridge, MA, 02138,USA}
 \affiliation{ $^{2}$
Clarendon Laboratory, Department of Physics, University of Oxford,
Parks Road, Oxford, OX1 3PU, UK} \affiliation{ $^{3}$ National
Institute of Standards and Technology, Gaithersburg MD, 20899, USA}
 \affiliation{$^{4}$ Dept. of Phys., George Mason U., Fairfax, VA, 22030,USA}
\date{\today}

\begin{abstract}
We study the effects of rotation on one-dimensional ultra-cold
bosons confined to a ring lattice. For commensurate systems, at a
critical value of the rotation frequency, an infinitesimal interatomic
interaction energy opens a gap in the excitation spectrum, fragments the
ground state into a macroscopic superposition of two states with
different circulation and generates a sudden change in the topology
of the momentum distribution. These features are reminiscent of the
topological changes in the Fermi surface that occurs in the Lifshitz
transition in fermionic systems. The entangled  nature of the ground
state induces a strong enhancement of quantum correlations and
decreases the threshold for the Mott insulator transition. In
contrast to the commensurate case, the incommensurate lattice is
rather insensitive to rotation. Our studies demonstrate the
utility of noise correlations as a tool for identifying new physics in
strongly correlated systems.

\end{abstract}
\maketitle

Ultra-cold gases loaded in an optical lattices are becoming one of
the most exciting platforms for exploring complex phases of strongly
correlated systems. Flexible experimental control over the parameters of the
lattice and the strength of the interatomic interactions have led to
the realization of fermionization of bosons, $i.e.$ the
Tonks-Girardeau (TG) regime \cite{Weiss,Paredes}, and a
one-dimensional Mott insulator (MI) state \cite{Paredes}.
 Recent experimental advances \cite{Helmerson}
 are opening  a new arena for the investigation of persistent
currents and various novel physics that emerge when cold atoms are
trapped in ring-shaped optical lattice. In these systems it is now
possible to create an artificial magnetic field by rotating the
atomic cloud \cite{Tung,Jaksch,Sørensen,Bhat} and to study  the
exotic phenomena that emerge when interacting particles are
subjected to strong gauge fields.

In this paper,  we investigate the effects of rotation on the ground
state properties of bosonic atoms confined in a one-dimensional ring
lattice. When the lattice is rotated with angular frequency
$\Omega$, the Coriolis force generates an effective vector
potential, and hence a net circulation or non zero vorticity.  The
physics of the interacting quantum many-body system differs
considerably between the commensurate case (CO:  where the number of
atoms, $N$, is an integral multiple of the number of lattice sites,
$L$) and the incommensurate case, ICO. Recent studies\cite{Keith}
have shown that at a critical frequency of rotation, $\Omega_c$, the
ground state of CO lattices becomes a macroscopic superposition (cat
state) of two states of opposite circulation \cite{Zhou}. Here we
focus on the role of the interatomic interaction. For CO lattices,
we show that a maximally entangled state exists at an infinitesimal
value of the on-site interaction parameter, accompanied by a
discontinuous rearrangement of momentum distribution, the opening of
a gap in the energy spectrum and non-analytic behavior of the
compressibility. These effects are analogous to the sudden change in
the topology of the Fermi surface in the conventional Lifshitz
transition  in fermionic systems \cite{lif}, in which a gap and a
discontinuous change in the momentum distribution  are induced as
the pressure approaches a critical value. We show that the entangled
nature of the ground state as the system approaches $\Omega_c$ is
accompanied by a large depletion of the condensate population, which
in turn drives the system to the MI phase  at lower interaction
strengths and by a strong enhancement of the intrinsic quantum noise
signaled  in the noise correlations. Our analysis demonstrates that
these higher order correlations, which can be measured
experimentally in time of flight images \cite{Altman, Foelling},
provide a unique experimental probe for detecting different quantum
phases in rotating lattices.

We consider  a system of $N$ ultra-cold bosons with
mass $M$ confined in a uniform 1D ring lattice of $L$ sites with  lattice constant $d$.  The ring is rotated about its axis  ($z$ axis)
with angular velocity $\Omega$ . In the rotating frame of the ring
the many-body Hamiltonian is given by

\begin{equation}
\hat {H}=\int d\mathbf{x}
\hat{\Phi}^{\dagger}\left[-\frac{\hbar^2}{2 M}\nabla^2 +
V(\mathbf{x}) +\frac{4 \pi \hbar^2 a}{2M}
\hat{\Phi}^{\dagger}\hat{\Phi}- \Omega \hat{L}_z\right]\hat{\Phi}
\end{equation} where $a$ is the $s$-wave
scattering length,$V(\mathbf{x})$ the lattice potential and
$\hat{L}_z$ the angular momentum. Assuming that the lattice is deep
enough to restrict tunneling between nearest-neighbor sites and the
band gap is larger than the rotational energy, the bosonic field
operator, $\hat{\Phi}$ can be expanded in Wannier orbitals confined
to the first band $\hat{\Phi}=\sum_j \hat{a}_jW_j'(\mathbf{x}) $,
$W_j'(\mathbf{x})=\exp\left[\frac{-i M}{\hbar}
\int_{\mathbf{x}_j'}^\mathbf{x} \mathbf{A}(\mathbf{x}')\cdot
d\mathbf{x}'\right]W_j(\mathbf{x})$. Here  $W_j(\mathbf{x})$ are the
Wannier orbitals of the stationary lattice centered at the site $j$,
$\mathbf{A}(\mathbf{x})=\Omega \hat{z} \times \mathbf{x}$ an
effective vector potential and $\hat{a}_j$ the bosonic annihilation
operator of a particle at site $j$ .   In terms of these quantities,
the many-body Hamiltonian can be written, up to onsite diagonal
terms which we neglect for simplicity, as \cite{Bhat,Wu}:

\begin{equation}
\hat{H}= -J\sum_{j}e^{-i \theta
}\hat{a}_j^{\dagger}\hat{a}_{j+1}+e^{i \theta
}\hat{a}_j\hat{a}_{j+1}^{\dagger}+
\frac{U}{2}\sum_{j}\hat{n}_j(\hat{n}_j-1) \label{ham}
\end{equation}Here
$\hat{n}_j=\hat{a}_j^{\dagger}\hat{a}_{j}$,  $\theta$ is the
effective phase twist  induced by the Gauge field, $\theta
=\int_{\mathbf{x}_{i}}^\mathbf{x_{i+1}} \mathbf{A}(\mathbf{x}')\cdot
d\mathbf{x}'=\frac{M \Omega L d^2}{h}$,  and $J$ is the hopping
energy:  $J \equiv\int d\mathbf{x} W_i^{*}\left[-\frac{\hbar^2}{2
M}\nabla^2 + V(\mathbf{x})\right]W_{i+1}$, and $U$ the on-site
interaction energy: $U\equiv \frac{4 \pi a \hbar^2}{M} \int
d\mathbf{x}|W_i|^4$ .

We first carry out a Bogoliubov analysis(BA) \cite{AnaJPB} to
understand the effects of rotation on the  bosonic ring in the
weakly interacting limit. The BA approximates the field operator
by a c-number plus small fluctuations $\hat{a}_{n}= z_{n}+\hat{%
\delta}_{n}$. Replacing it in the many-body Hamiltonian and
including terms up to second order in $\hat{\delta}$ yields a
quadratic Hamiltonian which can be diagonalized by the canonical
transformation, $\hat{\delta}_{n}=\sum_{q}u_{n}^{q}\hat{\alpha
}_{q}-v_{n}^{\ast q} \hat{\alpha }_{q}^{\dagger }$. For the
Hamiltonian described by Eq.(\ref{ham}) the
 diagonalization procedure yields: $z_n= e^{i \phi_k n}
\sqrt{n_0}$, $u^q_n= \frac{1}{\sqrt{L}} e^{i n (\phi_k+ \phi_q)}
u_q$ and $v^q_n= \frac{1}{\sqrt{L}} e^{i n (\phi_q-\phi_k)} v_q$,
with $\phi_k=\frac{2\pi}{L} k$, $k$ and $q$ integers and

\begin{eqnarray}
\mu&=&-2J\cos[\phi_k-\theta ]+n_0 U \label{eq1}\\
v_q^2&=&u_q^2-1=\frac{\epsilon_q+ U \bar{n}}{2(\sqrt{\epsilon_q^2+2
U \bar{n} \epsilon_q})}-\frac{1}{2} \label{eq2}
\end{eqnarray}where  $\mu $ is the chemical potential of the system and $\epsilon_q=4J \sin^2[\frac{\phi_q}{2}] \cos[\phi_k-\theta]$.
 $n_0=\bar{n}-\frac{1}{L}\sum_{q\neq 0}v_q^2 $ is the condensate density and $\bar{n}-n_0$
the condensate depletion.

If we write $\theta=\frac{2\pi}{L} m + \frac{ \Delta\theta}{L}$ with
$m$ an integer and  $0\leq \Delta\theta< 2\pi$,  the ground state
energy corresponds to $k=m$ if $ 0\leq \Delta\theta< \pi$ and
$k=m+1$ if $ \pi< \Delta\theta< 2\pi$. In order words,  to lower the
energy gained by rotation,  the condensate acquires a phase   and
becomes a current carrying state with quasi-momentum (QM)
$\frac{2\pi}{L} k$. The quantum number $k$ can also be viewed as a
measure of the circulation or vorticity of the gas. Therefore, a
change in $k$ from $m$ to $m+1$ at the critical angular velocity
$\Omega_c=\frac{\pi(2m+1) h}{M d^2 L^2}$ (i.e.  $\Delta
\theta=\pi$), implies the nucleation of a  vortex in the system. For
$\Omega\neq \Omega_c$ there is always a unique ground configuration
corresponding to a macroscopically  occupied state with QM equal to
$\frac{2\pi}{L} k$. On the contrary, at $\Omega_c$ the ground state
is not unique as the $k=m$ and $k=m+1$ solutions are degenerate. At
this critical rotation, we expect the BA to break down as the
underlying assumption of the existence of a single macroscopically
occupied mode is no longer valid. Explicit comparison between the BA and
numerical results, discussed later, shows that the BA provides a fair
description of the system for small $U$, and away from $\Omega_c$.

To understand the behavior of the system at critical rotation, we write
the many-body Hamiltonian in terms of QM operators,
$\hat{b}_q=\sum_{j=1}^{N}\hat{a}_j e^{-i 2\pi q j/L}$, $q=0,\dots, L-1$:

\begin{equation}
\hat{H}= \sum_{j=0 }^{N-1}
E_q\hat{b}_q^{\dagger}\hat{b}_{q}+\frac{U}{2L}\sum_{q,s,l=0}^{N-1}\hat{b}_q^{\dagger}
\hat{b}_s^{\dagger} \hat{b}_l\hat{b}_{\|q+s-l\|_L}
\end{equation} Here $E_q=-2J \cos[\phi_q-\theta] $ are  single-particle energies  and
the notation $\|\quad \|_L$ indicates modulo $L$. In the
non-interacting limit, at $\Omega_c$, all the $N+1$ states given by
$|n,N-n \rangle =\frac{1}{\sqrt{n!(N-n!)}} \hat{b}_m^{\dagger
n}\hat{b}_{m+1}^{\dagger (N-n)}|0\rangle$, with $n=0,\dots,N$ are
degenerate and span the ground state manifold. For small U, we
apply first order perturbation
theory by diagonalizing the Hamiltonian in the degenerate subspace. In view
of the absence of direct coupling between the different states, the
matrix is diagonal and the energy shifts are given by:
$E_{n}^{(1)}=E_m N +\frac{U}{2L}[(N+n)^2-3 n^2 -N]$. The
states with minimal energy are those corresponding to $n=0$ and $n=N$. To first
order in $U$, these two states remain degenerate and higher order
perturbation theory is needed to break the degeneracy. To understand
the effects of interaction on these degenerate states, one has
to distinguish between CO and ICO cases.

In the ICO case, it has been shown\cite{Keith} that there
is no coupling between $|N,0\rangle$ and $|0,N\rangle$ and hence
these states remain degenerate for all values of $U$.
Therefore, one can
take one of these states as the macroscopically occupied mode and
use the BA to take into account the depletion introduced by quantum
fluctuations as $U/J$ increases.

In contrast, in the  CO system, there are many different paths that
couple these two states and their number increases exponentially with
the number of atoms and the sites. The coupling leads to the opening
of a gap $\Delta E$. If we write an effective Hamiltonian between
the two states, we find that
\begin{equation}
V_{12} \equiv \langle 0,N|H_{eff}|N,0\rangle=\Delta
E=(\frac{U}{2L})^{\bar{n}(L-1)} A \label{gap}
\end{equation}with $A=\sum_{i,j,\dots p}
\frac{H_{0i} H_{ij}\dots H_{pN}}{(E_{0}^{(1)} -\varepsilon_i)
(E_{0}^{(1)} -\varepsilon_i) (E_{0}^{(1)}-\varepsilon_j) \dots (
E_0^(1) -\varepsilon_p)}$  being  $H_{ij}$, the transition  matrix
elements introduced by the interaction term of the Hamiltonian and
$\varepsilon_i$  the non-interacting many-body eigenenergies of the
intermediate states. The factor ${\bar{n}(L-1)}$ corresponds to the
minimum number of collision processes necessary to couple the states
$|N,0\rangle$ and $|0,N\rangle$  and the  sum is over  all the
different paths that generate such  couplings. The ground state of
the system becomes :
\begin{equation}|\Psi\rangle=\frac{a_1
|0,N\rangle+a_2|N,0\rangle}{\sqrt{2}} \label{cat}
\end{equation} with $\frac{a_1}{a_2}=-\frac{V_{12}}{|V_{12}|}$ (see
\cite{Keith}). Note that any infinitesimal value of  $U$ is enough
to generate the superposition. As the interaction increases, the
states corresponding to other QMs also contribute to the ground state
and at some point the system becomes a MI state. Therefore, at criticality
($\Omega=\Omega_c$), the ICO and CO respond very differently to
the rotation.

Fig. ~1 shows the variation in the spectral gap with interaction for
different values of rotation. The critical rotation is
characterized by the opening of a gap at small $U$ with a power-law
dependence on $U$ as predicted by Eq. \ref{gap}. Furthermore, the
compressibility ($d\mu/d\bar{n}$) shows a singular behavior at
integer fillings(see the inset in Fig.1).
Away from criticality, the system undergoes the usual Mott
transition at finite but reduced value of $U/J$. The decrease in the
MI threshold can be explained by means of  a mean field  analysis
and will be  discussed later in the paper.
\begin{figure}[htbp]
\includegraphics[width=2.8in]{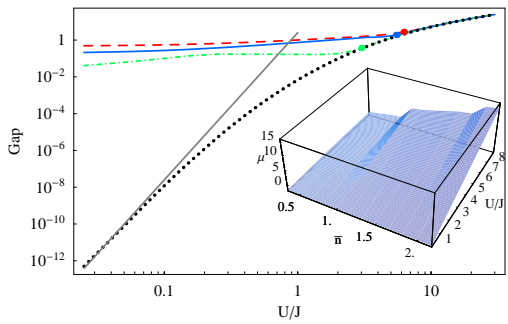}
\leavevmode \caption{ (color online)Gap for $L=9$ and $N=9$ as a
function of $U/J$. The solid (red), dashed (blue), dotted-dashed
(green) and dotted (black) curves are for $\Delta \theta/2\pi=0,
0.3,0.48$ and $0.5$ respectively. The grey line is  proportional to
$(U/J)^{\bar{n}(L-1)}$. The dots correspond to $U_c$ calculated from
the maximum in the noise correlations ( see Fig. ~3). The inset
shows the variation in the chemical potential ($\mu$) vs  density
($\bar{n}$) and $U/J$  at $\Omega_c$. } \label{fig1}
\end{figure}

The  complexity inherent to the highly entangled nature of the
ground  state  as $\Omega$ approaches $\Omega_c$, and the striking
different behavior between the CO and ICO system can be best
illustrated by  the first and the second order correlations, namely
the momentum distribution $\hat{n}(Q)=\frac{1}{L}\sum_{m,n} e^{i 2
 \pi Q (m-n) /L }\hat{a}_m^\dagger \hat{a}_n$,
$Q=0,1,\dots,L-1$ and the noise correlations,
$\Delta(Q_1,Q_2)= \langle\hat{n}(Q_1)
\hat{n}(Q_2)\rangle
-\langle\hat{n}(Q_1)\rangle\langle\hat{n}(Q_2)\rangle$.

\begin{figure}[htbp]
\includegraphics[width=3.5in]{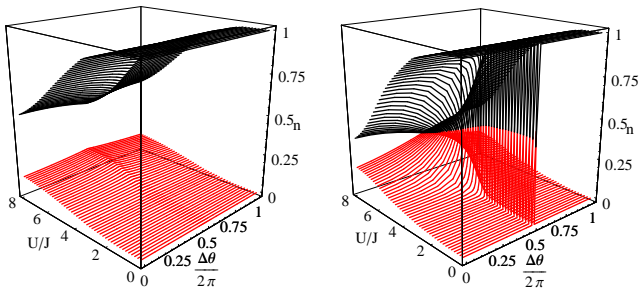}
\leavevmode \caption{ (color online)  n(Q)/L  vs. $U/J$ and $\Delta
\theta$ for $L=9$: Right $N=9$, Left $N=8$. Black: $Q=k$ (condensate
mode ), red:  $Q=k\pm1$ ($\pm$ depending if $\Delta
\theta\lessgtr\pi$) } \label{fig2}
\end{figure}


In Fig. ~2 we display the fractional number of atoms in the $Q=k$ and
$Q=k\pm1$ modes as a
function of $\Delta \theta$ and $U/J$,
with $\pm$ determined by $\Delta \theta\lessgtr\pi$.  In the ICO case there is a single macroscopically occupied mode even at
$\Omega_c$, and the
nature of the ground state is weakly dependent on $\theta$. In contrast, for the CO system there is an abrupt redistribution of
the population of atoms as $\theta $ approaches $\Omega_c$. At
criticality, for any  $U/J>0$, the population of the  $Q=k$ and
$Q=k\pm1$ modes becomes  identical. While for small U, only
these two modes are occupied,  other modes become populated as
$U$ increases and for $U/J \gg 1$, irrespective of the $\theta$
value, the state of the system becomes an equal superposition of all
the different QM components.

The sudden rearrangement of atoms in different QM states changes the
topology of the  $\hat{n}(Q=k,U,\theta)$ and $\hat{n}(Q=k\pm
1,U,\theta)$surfaces: they touch at the critical rotation frequency
illustrating the entanglement. This redistribution of momentum,
accompanied by an opening of a gap in the spectrum and a
non-analyticity in the compressibility ( see fig. ~1) is reminiscent
of the Lifshitz transition in fermionic system where a change in the
pressure beyond a critical value results in an abrupt change in the
Fermi surface. In other words, rotation mimics the effect of pressure
and induces a Lifshitz-like transition in bosonic atoms confined to
a ring shaped geometry.

The existence of maximal entanglement at the onset to the
transition can be effectively demonstrated by the noise correlations
which maps the intrinsic quantum noise into an intensity signal.
Away from the critical frequency, the onset of the superfluid to MI
transition at $U_c$ is signaled by the appearance of a maximum in
the intensity of $\Delta(k,k)$ \cite{reynoise}. However, at
$\Omega_c$ (see Fig.~3), the maximal entanglement
occurs at infinitesimal $U$, coinciding with the formation of the cat state.
 As a consequence the usual peak at  $U_c$ is transformed to an
abrupt jump in the second-order correlations at $U>0$. This jump is
correctly predicted by Eqs. (\ref{cat}), which yields $\hat{n}(Q)=
\frac{N}{2}( \delta_{m,Q} + \delta_{m+1,Q})$ and
$\Delta(m,m)=\Delta(m+1,m+1)=-\Delta(m,m+1)=\frac{N^2}{4}$ and zero
otherwise, in excellent agreement with the numerical simulations.
Fig.~ 3 also shows a faster convergence to the asymptotic behavior,
$\Delta(Q,k)\to 2\delta_{Q,k}$ and $n(Q)\to \bar{n}$, characteristic
of a system deep in the MI phase with suppressed number fluctuations
\cite{Altman,reynoise}.

\begin{figure}[htbp]
\includegraphics[width=3.5in,height= 2.0in]{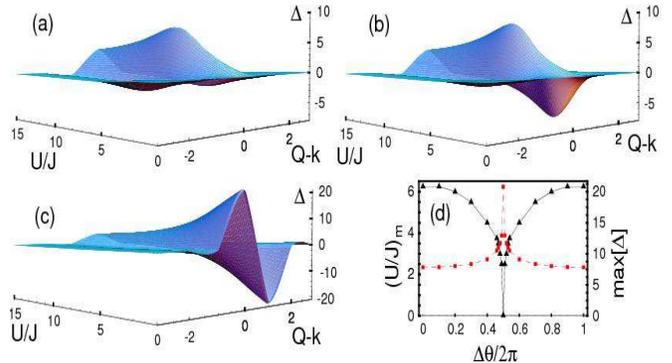}
\leavevmode \caption{ (color online) (a-c) shows $\Delta(Q-k,k)$ vs
$Q-k$ and $U/J$ for  $L=9$ and $N=9$ and  $\theta/L=0, 0.4, 0.5$.
The plots show that $\Delta(Q-k,k)$ exhibits  a maximum at the onset
of the MI for $\theta < \theta_c$. At $\theta_c$ a jump at $U>0$
occurs. In the inset (d) we plot the intensity of the maximum
intensity reached by $\Delta(k,k)$ (triangles right scale) and the
$U/J$ value at which it takes place (boxes left scale)
 } \label{fig3}
\end{figure}

The  inset of Fig.~3 illustrates the enhancement of quantum
correlations as the system approaches $\Omega_c$.  For a
given $\Delta \theta$,  we plot the maximum value of $\Delta(k,k)$
as $U/J$ is varied. Furthermore, we also show the actual $U/J$
value at which the maximum is reached. As stated earlier, this
value for $\Omega\neq \Omega_c$ is related to the onset of the MI
transition, and so the plot demonstrates a shift in $U_c$  as the
the rotation approaches the critical frequency.
The Gutzwiller ansatz (GA)\cite{Sheshadri} provides further
insight into the MI transition in the rotating ring lattice.
The  generic GA is based on the assumption that the wave
function can be approximated as a product of the wave functions at
the various lattice sites,$ |\varphi
_{G}\rangle=\prod_{i=0}^{N}\left(\sum_{n=0}^{\infty}f^{(i)}_{n}\left|n
\right\rangle_{i}\right)$. Here $|n\rangle_{i}$ is a Fock states
with n atoms at site $i$. Deep in the MI phase, the only relevant states are those
characterized by
$\bar{n}$, $\bar{n}+1$ and $\bar{n}-1$ atoms per site.
Assuming
$f^{(j)}_{\bar{n}}=\sqrt{\frac{1-\epsilon}{L}}e^{i \phi_k j}$,
$f^{(j)}_{\bar{n}+1}=f^{(j)}_{\bar{n}-1}=\sqrt{\frac{\epsilon }{2
L}} e^{i \phi_k j}$ and $f^{(j)}_{n\neq \bar{n}, \bar{n}\pm 1}=0$,
and minimizing the energy  $E=\langle\varphi_
{G}|\hat{H}|\varphi_{G}\rangle$ with respect to
$\epsilon$ and $\phi_k$, $\epsilon$ is found to vanish at
 $U_c$ given by:
 \begin{equation}
U_c=2J \cos(2 \pi \Delta \theta/L)
(\sqrt{\bar{n}}+\sqrt{\bar{n}+1})^2 \label{uc}\end{equation} where
$\Delta \theta=2\pi k/L$ with $k=m$ for $0\leq \Delta \theta< \pi$
and   $k=m+1$  for $\pi <\Delta \theta <2\pi$. Thus, the GA predicts a degenerate ground state at $\Omega_c$, in
agreement with BA.
Furthermore, it also shows a decrease of the critical value of the
MI transition threshold with increasing  rotation by a factor of
$\cos(2 \pi \Delta \theta/L)$. The shift is much smaller than that
value found in our numerical calculations, but this is expected
because mean field theories only provide a qualitative analysis in
1D systems.

For the ICO system, correlations can be computed analytically in both
small and the large $U$ limit. For small $U$, using Eq. (\ref{eq1})
and (\ref{eq2}), we can determine the $n(Q)$ and
$\Delta(Q,Q')$ . The resulting momentum distribution $\hat{n}(Q)=L n_0
\delta_{k,Q} + v_{k-Q}^2$ implies a sharp interference peak at the
QM of the condensate whose width increases with interactions due to
quantum depletion. On the other hand, noise correlations are
directly related to the quantity $u_q v_q$, the so called anomalous
average \cite{Hutchinson} which  is  a measure of the many-body
scattering matrix and  accounts for the modification of binary
scattering properties due to the presence of other surrounding
atoms. The only non-vanishing correlations are $\Delta(k,k)$,
$\Delta(k,Q)$, $\Delta(Q,Q)$ and  $\Delta(2k-Q,Q)$ with $Q \neq k$.
$\Delta(k,Q)=-2u_{k-Q}^2 v_{k-Q}^2$ probes  pair excitations of the
condensate and   has a negative intensity. The anti-correlation  can
be understood as a consequence  of the destructive interference
between the excited  pairs generated when two condensate atoms
collide. $ \Delta(Q,Q)=\Delta(Q,2k-Q)=u_{k-Q}^2 v_{k-Q}^2$ probe
 collisions between atoms out-of the condensate; these processes
are less frequent than condensate collisions  and that is why they
 have half of the   intensity of  $\Delta(k,Q)$.
$\Delta(k,k)=2 \sum_{q\neq 0}u_q^2 v_q^2$, has the larger intensity
as its value   is twice the sum of the intensities of the other
$\Delta(Q,Q)$ peaks. Fig. ~4 shows a comparison between analytic
results obtained within BA approximation and the exact numerical
diagonalization of Eq. (2) for a finite ICO system. We plot
$\Delta(m,m)$, $\Delta(m+1,m+1)$ and $\Delta(m,m+1)$ as a function
of the ratio $U/J$ for different rotation velocities.  The agreement
in the small $U$ for a system of only $9$-lattice sites is
reassuring that our analysis based on a finite lattice with few
atoms captures relevant features.

\begin{figure}[htbp]
\includegraphics[width=3.5in]{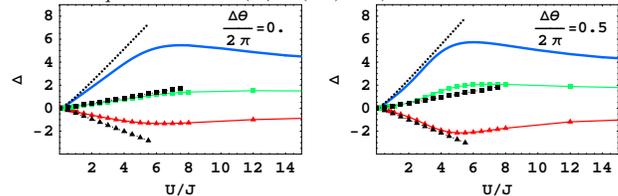}
\leavevmode \caption{(color online) Noise correlation vs. $U/J$ for
$L=9$ and $N=8$ : Exact (solid lines),  BA  ( black lines). We use
no symbols for $\Delta(m,m)$, triangles for $\Delta(m,m+1)$and boxes
for $\Delta(m+1,m+1)$. } \label{fig4}
\end{figure}

In the  strongly interacting limit, the behavior of the system
can be described by  mapping the  repulsive bosons into a gas
of ideal fermions \cite{GR}. For simplicity we will assume
$\bar{n}<1$, although the
mapping to fermions  can be generalized to larger filling factors
 \cite{Guido}). In this limit, the single-particle  energies are given
by $E_q=-2J\cos(\pi/L(2q-1) -\theta)$ when $N$ is even, and
$E_q=-2J\cos(\pi/L(2q) -\theta)$ for odd $N$ \footnote{The different
energy spectrum for even and odd number of particles comes from the
fact that when N is even the mapping to fermions requires to use
anti-periodic boundary condition}. These equations clearly show that
for an ICO system  at $\Omega_c$ there is always one empty
energy level with the same energy as the Fermi level. Therefore,
the two-fold degeneracy of the ICO ground state can be seen
even in the $U \to \infty$ limit.


In summary, our study demonstrates the effectiveness of
experimentally-accessible noise correlations in capturing the
complexity of highly entangled states, and in heralding a novel type
of phase transition. Our analysis also shows the appeal of noise
correlation spectroscopy  as a probe of  the onset of the  MI
transition in finite systems, where the condensate population
decreases gradually and thus the momentum distribution does not
provide a distinctive signature. Our studies raise important open
questions regarding the onset and nature of the MI at $\Omega_c$ as
instead of the standard transition from a superfluid to a MI here we
have a transition from a gapped, fragmented and incompressible state
to a MI. We hope that our suggestion of Lifshitz-like transition in
bosonic system will stimulate a new line of research in condensed
matter community.

{\it{Acknowledgments}} We are grateful to Carlos Sa de Melo and
David Hallwood for very stimulating and useful discussions and
suggestions. A.M.R. acknowledges support by a grant from the
Institute of Theoretical, Atomic, Molecular  and Optical Physics at
Harvard University and Smithsonian Astrophysical observatory.

\end{document}